\documentstyle[11pt]{article}

\def\theequation{\arabic{section}.\arabic{equation}}

\renewcommand{\theequation}{\thesection.\arabic{equation}}

\global\arraycolsep=1pt
\input epsf
\input{tcilatex}
\begin{document}

\font\cmss=cmss10 \font\cmsss=cmss10 at 7pt \hfill \hfill IFUP-TH/01-11


\vspace{10pt}

\begin{center}
{\Large {\bf \vspace{10pt} KINEMATIC SUM RULES FOR TRACE ANOMALIES }} 
\vspace{10pt}

\bigskip \bigskip

{\sl D. Anselmi}

{\it Dipartimento di Fisica, Universit\`a di Pisa, via F. Buonarroti 2,
56126 Pisa, Italia}
\end{center}

\vskip 2truecm

\begin{center}
{\bf Abstract}
\end{center}

I derive a procedure to generate sum rules for the trace anomalies $a$ and $%
a^{\prime }$. Linear combinations of $\Delta a\equiv a_{{\rm UV}}-a_{{\rm IR}%
}$ and $\Delta a^{\prime }\equiv a_{{\rm UV}}^{\prime }-a_{{\rm IR}}^{\prime
}$ are expressed as multiple flow integrals of the two-, three- and
four-point functions of the trace of the stress tensor. Eliminating $\Delta
a^{\prime }$, universal flow invariants are obtained, in particular sum
rules for $\Delta a$. The formulas hold in the most general renormalizable
quantum field theory (unitary or not), interpolating between UV and IR
conformal fixed points. I discuss the relevance of these sum rules for the
issue of the irreversibility of the RG flow. The procedure can be
generalized to derive sum rules for the trace anomaly $c$.

\vspace{4pt}

\vskip 1truecm

\vfill\eject

\section{Introduction}

\setcounter{equation}{0}

Quantum field theory of particles and fields of spin 0, 1/2 and 1 can be,
without loss of generality, embedded in external gravity. The gravitational
embedding can be useful to study properties of the ultraviolet and infrared
fixed points of the renormalization-group (RG) flow. The correlation
functions of the stress-tensor $T_{\mu \nu }$ are encoded in the induced
action for the gravitational background and define quantities which
characterize conformal and running quantum field theories.

At criticality, the trace anomaly in external gravity defines two central
charges, denoted by $c$ and $a$: 
\begin{equation}
\Theta _{*}={\frac{1}{(4\pi )^{2}}}\left[ c\,W^{2}-{\frac{a}{4}}\,{\rm G}+{%
\frac{2}{3}}\,a^{\prime }\,\Box R\right] ,  \label{crit}
\end{equation}
where $W$ is the Weyl tensor and ${\rm G}=4R_{\mu \nu \rho \sigma }R^{\mu
\nu \rho \sigma }-16R_{\mu \nu }R^{\mu \nu }+4R^{2}$ is the Euler density.
In free-field theories, we have $c=(n_{s}+6n_{f}+12n_{v})/120$ and $%
a=(n_{s}+11n_{f}+62n_{v})/360$, where $n_{s}$, $n_{f}$ and $n_{v}$ are the
numbers of real scalar fields, Dirac fermions and vectors, respectively. The
quantity $a^{\prime }$ is more peculiar, because it does not have a definite
value at criticality.

Off-criticality, $c$ and $a$ depend on the energy scale. In a variety of
cases, it is possible to compute the exact IR values of $c$ and $a$ in
asymptotically-free supersymmetric gauge theories \cite{noi,noi2}.

The exact results of \cite{noi,noi2} show that the UV values of the central
charge $a$ is always larger than its IR value, as conjectured by Cardy in 
\cite{cardy}. The property $a_{{\rm UV}}\geq a_{{\rm IR}}$ is named
``irreversibility of the RG flow'', or ``$a$-theorem''. Sometimes, the name
``$c$-theorem'' is borrowed from the two-dimensional theorem of
Zamolodchikov \cite{zamolo}. The central charge $c$ does not satisfy an
analogous property in four dimensions.

The study of the gravitational embedding in four dimensions is in general a
difficult task. The perturbative calculations have been pioneered by
Hathrell \cite{hathrell,hathrell2}, who worked out the values of $c$, $a$,
and $a^{\prime }$ to the second ($c$) and third ($a,a^{\prime }$) loop
orders. More general methods combine conformal properties and
renormalization-group techniques \cite{central}. Recently, Cappelli {\it et
al.} \cite{cappelli} have classified the structure of the stress-tensor
three-point function off criticality and obtained sum rules for the
anomalies $c$ and $a$. This classification is rather involved. Conceivably,
the classification of the four-point function, which is potentially useful
for the investigation of the irreversibility of the RG flow, is even harder.
Another approach to the sum rules for trace anomalies is the one of \cite
{hathrell,hathrell2} and \cite{jack}.

Other important results concern the induced action $\Gamma $ for the
gravitational background. Riegert \cite{riegert} and others \cite
{townsend,antoniadis} integrated the critical expression (\ref{crit}) of the
trace anomaly with respect to the conformal factor. This procedure gives $%
\Gamma $ up to conformally invariant terms. The conformally invariant terms
missed by this method have not been written in closed form, so far. The
Riegert action is made of some non-local terms, containing $c$ and $a$, plus
a unique, arbitrary, local term, $\int \sqrt{g}R^{2}$, multiplied by $%
a^{\prime }$. The locality of this term explains why $a^{\prime }$ has no
definite value at criticality and can be shifted by an arbitrary constant.
This shift does not depend on the energy and disappears in the difference
between the values of $a^{\prime }$ at two energies, e.g. $\Delta a^{\prime
}\equiv a_{{\rm UV}}^{\prime }-a_{{\rm IR}}^{\prime }$. Nevertheless, $%
\Delta a^{\prime }$ remains dependent on the flow connecting the two fixed
points, as shown in \cite{inv}.

A considerable simplification occurs in conformally flat backgrounds. The
correlation functions containing an arbitrary number of insertions of the
trace $\Theta $ of the stress tensor can be studied. The restricted
embedding looses track of the central charge $c$, but keeps track of $a$.
The Riegert action specialised to conformally flat metrics is local and
complete. It contains only two independent terms, multiplied by $a$ and $%
a^{\prime }$.

The so-specialised Riegert action encodes the UV and IR expressions of the $%
\Theta $-correlators, in terms of $a$ and $a^{\prime }$. In this paper, I
derive sum rules for the trace anomalies in the most general renormalizable
(not necessarily unitary) quantum field theory interpolating between UV and
IR conformal fixed points. The formulas are obtained exploiting the fact
that the $\Theta $-correlators have to tend to the UV\ and IR limits encoded
in $\Gamma $. These sum rules are called ``kinematic''. The procedure
naturally extends to more general background metrics, to derive sum rules
for the trace anomaly $c$. I\ study the conformally flat background in
detail and briefly describe this generalization.

The sum rules consist of flow integrals of the $\Theta $-correlators in
coordinate space, multiplied by polynomials of degree four in the
coordinates. Every flow integral is equal to a linear combination of $\Delta
a$ and $\Delta a^{\prime }$. Combining the sum rules, it is possible to
eliminate $\Delta a^{\prime }$ and obtain flow invariants, in particular sum
rules for $\Delta a$. A flow invariant is a (multiple) flow integral of a
correlator, whose value depends only on the extrema of the flow. Checks of
the sum rules in massive theories are presented in detail. Further
calculations, performed recently \cite{nuovo,fest}, show explicitly that the
flow-dependence of $\Delta a^{\prime }$ cancels out in the sum rules for $%
\Delta a$, which therefore appear to be meaningful.

I discuss the possible applications of the sum rules to the problem of the
irreversibility of the RG flow, comparing different, equivalent sum rules
for $\Delta a$. These do not appear to imply the irreversibility of the RG
flow in a straightforward way. In particular, I discuss certain difficulties
to apply Osterwalder-Schrader positivity \cite{oster}.

Finally, I describe the meaning of the results of \cite{athm} in the new
framework. In \cite{athm}, it was shown that a physical principle, suggested
by the properties of renormalization, implies a certain sum rule for $\Delta
a$ in unitary, classically conformal theories. The formula of \cite{athm}
involves only the $\Theta $-two-point function. The idea of \cite{athm} can
be collected into a ``dynamical'' vanishing sum rule for the $\Theta $%
-four-point function, not contained in the set of ``kinematic'' sum rules
worked out here.

The approach of this paper can be considered alternative, if not competing,
with those of \cite{cappelli} and \cite{hathrell,hathrell2,jack}. At the
moment, it is not clear which approach is more convenient for practical
computations.

The construction of this paper generalises to arbitrary even dimensions \cite
{fest}.

The paper is organised as follows. In sect. 2, I introduce the notation and
the general framework for the gravitational embedding and discuss the
convergence of the flow integrals. In sect. 3, I study the UV\ and IR\
limits of the $\Theta $-correlators\ and derive the sum rules. I comment on
the possible scheme dependence of certain flow integrals and the scheme
independence of the sum rules. In sect. 4, I derive flow invariants from the
sum rules. In sect. 5, I give explicit examples and checks of the formulas.
In sect. 6, I\ show that most sum rules are consequences of simple algebraic
symmetry properties of the integrals, plus the property that an
integrated-trace insertion is a scale derivative. In sect. 7, I write
explicit sum rules for $\Delta a$, focusing, in particular, on unitary,
classically conformal quantum field theories. I discuss the relevance of
these formulas for the issue of the irreversibility of the RG flow. I also
comment on the relation between the results of the present paper and those
of \cite{athm} and on the difficulties to apply OS positivity. Section 8
contains the conclusions.

\section{Preliminaries}

\setcounter{equation}{0}

In this section I describe the gravitational embedding and the
regularization technique. I also comment on the convergence of correlators.

{\bf Gravitational embedding.} The embedded theory is renormalizable \cite
{hathrell,hathrell2}. I add bare lagrangian terms of the form $\Gamma
_{0}[g_{\mu \nu }]=\Gamma _{a}[g_{\mu \nu }]+\Gamma _{b}[g_{\mu \nu }]$,
where 
\begin{equation}
\Gamma _{a}[g_{\mu \nu }]=\int {\rm d}^{n}x~\sqrt{g}\left[ a_{0}W^{2}+b_{0}%
{\rm G}+c_{0}R^{2}\right] ,\qquad \Gamma _{b}[g_{\mu \nu }]~=\int {\rm d}%
^{n}x~\sqrt{g}\left[ M_{0}^{2}R+\Lambda _{0}\right] .  \label{bobo}
\end{equation}
I use the dimensional-regularization technique in the Euclidean framework.
The space-time dimension is $n=4-\varepsilon $. A further term, $\int {\rm d}%
^{n}x\,\sqrt{g}\Box R$, can be omitted in $\Gamma _{a}$, since $\sqrt{g}\Box
R$ is a total derivative in every $n$. The integral $\int {\rm d}^{n}x\,%
\sqrt{g}{\rm G}$ is kept, since ${\rm G}$ reduces to the Euler density only
in four dimensions. The expression of ${\rm G}$ in $n$ dimensions is equal
to the one given in the previous section. The integral $\int {\rm d}^{n}x\,%
\sqrt{g}W^{2}$ is conformal invariant only in four dimensions. The
coefficients $a_{0},b_{0},c_{0},M_{0}$ and $\Lambda _{0}$ are independent
(bare) parameters, which appropriately reabsorb the divergences. In
classically conformal theories (e.g. massless QED), $M_{0}$ and $\Lambda
_{0} $ are absent \cite{hathrell,hathrell2}. I\ have separated the
``dimensionless divergences'' $\Gamma _{a}[g_{\mu \nu }]$ from the
``dimensioned divergences'' $\Gamma _{b}[g_{\mu \nu }]$ for later
convenience.

The induced action for the external metric is defined as 
\begin{equation}
\Gamma [g_{\mu \nu }]\equiv \Gamma _{0}[g_{\mu \nu }]~+\Gamma ^{\prime
}[g_{\mu \nu }]=\Gamma _{0}[g_{\mu \nu }]~{\rm -\ln }\int [{\rm d}\varphi ]~%
{\rm \exp }\left( -S[\varphi ,g_{\mu \nu }]\right) ,  \label{prime}
\end{equation}
where $\varphi $ collectively denotes the dynamical fields of the theory and 
$S[\varphi ,g_{\mu \nu }]$ is the action embedded in the external metric $%
g_{\mu \nu }$.

A great simplification occurs, if the background metric is restricted to be
conformally flat, $g_{\mu \nu }=\delta _{\mu \nu }~{\rm e}^{2\phi }$. The
conformal factor $\phi$ couples to the trace $\Theta $ of the stress tensor.
It can be proved that the $\Gamma _{a}[\phi ]$ is finite. Its $\varepsilon
\rightarrow 0$ limit is precisely the UV expression of the Riegert action
for conformally flat metrics.

More precisely, we can distinguish two classes of theories: the
classically conformal theories and the theories which are not conformal at
the classical level.

In classically conformal theories, the quartic, cubic, quadratic and linear
diverges can be canonically set to zero. This means that $\Gamma_b[\phi]$ is
absent. Moreover, it can be proved that the logarithmic divergences of the $%
\Theta $-correlators resum to zero. This convergence property is known for
the two-point correlator of $\Theta $ (see for example \cite{athm} and sect.
1.1 of \cite{inv}). In this paper, it will be proved in complete generality
(it is a consequence of the convergence of the sum rules). 
The convergence of the sum rules proves also that the action 
$\Gamma[\phi]$ is fully
convergent off criticality in the classically conformal theories.

If the theory is not conformal at the classical level (e.g. some field are
massive), $\Gamma _{b}[\phi ]$ remains divergent. Nevertheless, this term
does not affect the sum rules. It can be projected away imposing certain
restrictions on the test functions. The argument mentioned above to show
that $\Gamma _{a}[\phi ]$ has a finite $\varepsilon\rightarrow 0$ limit
applies also to this case.

The induced action for the conformal factor $\phi $ (the $\varepsilon
\rightarrow 0$ limit of $\Gamma _{a}[\phi ]$) reads at criticality \cite
{riegert,townsend,antoniadis} 
\begin{equation}
\Gamma ^{*}[\phi ]={\frac{1}{8\pi ^{2}}}\int {\rm d}^{4}x\,\left\{
a_{*}(\Box \phi )^{2}+(a_{*}^{\prime }-a_{*})\left[ \Box \phi +(\partial
_{\mu }\phi )^{2}\right] ^{2}\right\} .  \label{rest}
\end{equation}
This is the specialisation of the Riegert action \cite{riegert} to
conformally flat metrics (see also \cite{athm}).

I stress that I specialise {\sl ab initio} to conformally flat metrics $%
\delta _{\mu \nu }{\rm e}^{2\phi }$, and use the dimensional-regularization
technique. This strategy is convenient for the purposes of this paper, but
has a little drawback:\ since $\Gamma ^{*}[\phi ]$ is local, the critical
values of the coefficient $a$ are not calculable in this framework. The
coefficient $a $ can be calculated at criticality in the following two
cases: when the dimensional-regularization technique is used, but the
background metric $g_{\mu \nu }$ is kept generic -- then $a$ multiplies a
non-local term of the stress-tensor three-point function and a pole of $%
\Gamma _{a}[g_{\mu \nu }]$; when a different regularization technique (e.g.
Pauli-Villars) is used -- then it is possible to use directly the
conformally flat metric. In practice, in the framework of this paper, $a$
behaves like $a^{\prime }$, since both terms of $\Gamma ^{*}[\phi ]$ are on
the same footing. This has no effect on the calculations of this paper,
which are about the differences $\Delta a=a_{{\rm UV}}-a_{{\rm IR}}$ and $%
\Delta a^{\prime }=a_{{\rm UV}}^{\prime }-a_{{\rm IR}}^{\prime }.$

Summarising, we can write 
\[
\Gamma [\phi ]=\Gamma ^{*}[\phi ]+\Gamma _{b}[\phi ]{\rm -\ln }\int [{\rm d}%
\varphi ]~{\rm \exp }\left( -S[\varphi ,\phi ]\right) . 
\]
It will be proved later that this $\Gamma ^{*}[\phi ]$ is precisely $\Gamma
^{{\rm UV}}[\phi ]$, which means expression (\ref{rest}) with $%
a_{*},a_{*}^{\prime }\rightarrow a_{{\rm UV}},a_{{\rm UV}}^{\prime }$.

\section{Derivation of the sum rules}

\setcounter{equation}{0}

In this section, I derive the sum rules for $\Theta $-correlators.

Let $T_{\mu \nu }=2/\sqrt{g}~\delta S/\delta g^{\mu \nu }$ be the stress
tensor, $\Theta $ its trace and $\overline{\Theta }=\sqrt{g}\Theta= -\delta
S/\delta\phi$ . The functional derivatives $\delta ^{(k)}\Gamma [\phi
]/(\delta \phi (x_{1})\cdots \delta \phi (x_{k}))$ of the induced action $%
\Gamma$ restricted to the conformal factor $\phi$ are denoted by $\Gamma
_{x_{1}\cdots x_{k}}$. A similar notation is used for $\Gamma _{x_{1}\cdots
x_{k}}^{\prime }$ and the functional derivatives of $\overline{\Theta }(x)$.

I begin with the relations between the functional derivatives of $\Gamma
^{\prime }$ and the $\overline{\Theta }$-correlators. Successively
differentiating, we have $\Gamma _{x}^{\prime }=-\langle \overline{\Theta }%
(x)\rangle $ and 
\begin{eqnarray}
\Gamma _{x_{1}x_{2}}^{\prime }[\phi ] &=&-\langle \overline{\Theta }(x_{1})\,%
\overline{\Theta }(x_{2})\rangle -\langle \overline{\Theta }%
_{x_{2}}(x_{1})\rangle ,  \nonumber \\
\Gamma _{x_{1}x_{2}x_{3}}^{\prime }[\phi ] &=&-\langle \overline{\Theta }%
(x_{1})\,\overline{\Theta }(x_{2})\,\overline{\Theta }(x_{3})\rangle
-\langle \overline{\Theta }_{x_{3}}(x_{1})\,\overline{\Theta }(x_{2})\rangle
-\langle \overline{\Theta }(x_{1})\,\overline{\Theta }_{x_{3}}(x_{2})\rangle
\nonumber \\
&&~~~~~~~~~~~~~~~~~~~~~~~~~~~-\langle \overline{\Theta }_{x_{2}}(x_{1})\,%
\overline{\Theta }(x_{3})\rangle -\langle \overline{\Theta }%
_{x_{2}x_{3}}(x_{1})\rangle ,  \nonumber \\
\Gamma _{x_{1}x_{2}x_{3}x_{4}}^{\prime }[\phi ] &=&-\langle \overline{\Theta 
}(x_{1})\,\overline{\Theta }(x_{2})\,\overline{\Theta }(x_{3})\,\overline{%
\Theta }(x_{4})\rangle -\sum_{\{6\}}\langle \overline{\Theta }%
_{x_{l}}(x_{i})\,\overline{\Theta }(x_{j})\,\overline{\Theta }(x_{k})\rangle
-\sum_{\{4\}}\langle \overline{\Theta }_{x_{k}x_{l}}(x_{i})\,\overline{%
\Theta }(x_{j})\,\rangle  \nonumber \\
&&~~~~~~~~~~~~~~~~~~~~~~~~~~-\sum_{\{3\}}\langle \overline{\Theta }%
_{x_{k}}(x_{i})\,\overline{\Theta }_{x_{l}}(x_{j})\,\rangle -\langle \,%
\overline{\Theta }_{x_{2}x_{3}x_{4}}(x_{1})\rangle ,  \label{none}
\end{eqnarray}
etc. The notation is as follows. In the expression $\overline{\Theta }%
_{x_{j}x_{k}\cdots x_{l}}(x_{i})$, it is understood that $i<j,k,\cdots ,l$.
The number in curly brackets is the number of ways to distribute the indices
with this constraint. The symbol $\langle \cdots \rangle $ denotes the
connected components of the correlators.

\medskip

\medskip {\bf Criticality.} Using (\ref{rest}), we find that, at
criticality, $\Gamma _{x}^{*}[0]=0$ and $\Gamma _{x_{1}\cdots
x_{k}}^{*}[\phi ]=0$ for $k>4$. With the help of test functions $u$, we find
that the other functional derivatives of $\Gamma $ satisfy 
\begin{eqnarray}
\int {\rm d}^{4}x~u(x)~\Gamma _{x0}^{*}[0] &=&\frac{1}{4\pi ^{2}}%
a_{*}^{\prime }\Box ^{2}u(0),~~~~~~~~  \nonumber \\
\int {\rm d}^{4}x~{\rm d}^{4}y~u(x,y)~\Gamma _{xy0}^{*}[0] &=&\frac{%
a_{*}-a_{*}^{\prime }}{\pi ^{2}}\left[ \Box ^{x}\Box ^{y}-\left( \partial
^{x}\cdot \partial ^{y}\right) ^{2}\right] ~u(0,0),~~~~~~~~  \label{pred} \\
\int {\rm d}^{4}x\,{\rm d}^{4}y\,{\rm d}^{4}z\,u(x,y,z)\,\Gamma
_{xyz0}^{*}[0] &=&\frac{a_{*}-a_{*}^{\prime }}{\pi ^{2}}\left[ \Box
^{x}\partial ^{y}\cdot \partial ^{z}+2\,\partial ^{x}\cdot \partial
^{y}\,\partial ^{x}\cdot \partial ^{z}+{\rm cycl.perms.}\right] u(0,0,0). 
\nonumber
\end{eqnarray}
The last argument of $\Gamma _{x_{1}\cdots x_{k}}^{*}$ is set to zero using
translational invariance.

\medskip

{\bf Off-criticality.} Off-criticality, the correlators (or combinations of
correlators) $\Gamma _{x_{1}\cdots x_{k}}$ depend on the energy scale. We
want to study the UV and IR limits of $\Gamma _{x_{1}\cdots x_{k}}$ and
relate them to (\ref{pred}). Using suitable test functions (satisfying some
restrictions explained below), we have 
\begin{equation}
\lim_{{\rm UV\,(IR)}}\int \prod_{i=1}^{k}{\rm d}^{4}x_{i}~u(x_{1},\cdots
,x_{k})~\Gamma _{x_{1}\cdots x_{k}0}=\int \prod_{i=1}^{k}{\rm d}%
^{4}x_{i}~u(x_{1},\cdots ,x_{k})~\Gamma _{x_{1}\cdots x_{k}0}^{{\rm UV\,(IR)}%
}.  \label{eqw}
\end{equation}
It is understood that, after taking the $\phi $-derivatives of $\Gamma $, $%
\phi $ is set to zero.

The UV\ and IR\ limits are defined as follows. Let $\mu $ be the dynamical
scale. I collectively denote 
the dimensionful parameters of the theory with $m$. 
Concretely,\ in QCD we can take $\Lambda _{{\rm QCD}}$ and the
quark masses (or an equivalent number of independent hadron masses). After
the replacements $\mu ,m\rightarrow \lambda \mu ,\lambda m$ in $\Gamma
_{x_{1}\cdots x_{n}}$, the UV\ and IR\ limits are $\lambda \rightarrow 0$
and $\lambda \rightarrow \infty $, respectively.

The terms of the form $\Gamma _{b}[\phi ]$ contained in $\Gamma [\phi ]$ can
be projected away with a clever choice of the test functions $u$. The limits
(\ref{eqw}) exist if the theory interpolates between well-defined IR\ and
UV\ conformal fixed points, which I\ assume. The difference between the UV\
and IR\ values of the central charges $a$ and $a^{\prime }$ can then be
expressed by certain integrals. These are assured to be convergent by the
very same existence and finiteness of $\Gamma [\phi ]$ at criticality, where
it is equal to (\ref{rest}).

\medskip

{\bf Sum-rule generator.} I rescale $\mu $ and $m$ by a factor $\lambda $
and denote the rescaled correlators by $\Gamma _{x_{1}\cdots x_{k}}^{\lambda
}$. Making a change of variables $x_i\rightarrow x_i/\lambda$ in the
integrals, we get expressions of the form 
\[
\int \prod_{i=1}^{k}{\rm d}^{4}x_{i}~u(x_{1},\cdots ,x_{k})~\Gamma
_{x_{1}\cdots x_{k}0}^{\lambda }=\lambda ^{4}\int \prod_{i=1}^{k}{\rm d}%
^{4}x_{i}~u(x_{1}/\lambda ,\cdots ,x_{k}/\lambda )~\Gamma _{x_{1}\cdots
x_{k}0}. 
\]
As anticipated above, we have to impose certain conditions on the test
functions $u$. At $\{x_{i}\}=\{0\}$, we demand that $u$ vanishes together
with its first three derivatives with respect to all coordinates: 
\begin{equation}
u(0)=\partial _{i}u(0)=\partial _{i}\partial _{j}u(0)=\partial _{i}\partial
_{j}\partial _{k}u(0)=0,  \label{hyp}
\end{equation}
where $\partial_i=\partial/\partial x_i$. These conditions cancel out the
quartic, cubic, quadratic and linear divergences from the integrals, and
project $\Gamma _{b}[\phi ]$ away. From now on, I\ will always omit the
irrelevant terms $\Gamma _{b}[\phi ]$.

To implement the above conditions more explicitly, I take a test function of
the form 
\begin{equation}
u(x_{1},\cdots ,x_{k})=\sum_{\{k_{i}\}}\prod_{i=1}^{4}\frac{1}{k_{i}!}\left(
x_{i}\cdot \partial _{i}\right) ^{k_{i}}U(x_{1},\cdots ,x_{k}),  \label{para}
\end{equation}
where the sum runs over all sets of $k_{i}=0,1,2,3$ or 4, such that $%
\sum_{i=1}^{k}k_{i}=4$. The fourth derivatives of $u$ in the origin coincide
with those of $U$: $\partial _{i}\partial _{j}\partial _{k}\partial
_{l}u(0)=\partial _{i}\partial _{j}\partial _{k}\partial _{l}U(0)$. With the
parametrisation (\ref{para}), it is sufficient to demand that the function $%
U $ be regular and bounded. We have in the end 
\[
\int \prod_{i=1}^{k}{\rm d}^{4}x_{i}~\sum_{\{k_{i}\}}\prod_{i=1}^{4}\frac{1}{%
k_{i}!}\left( x_{i}\cdot \partial _{i}\right) ^{k_{i}}U(x_{1}/\lambda
,\cdots ,x_{k}/\lambda )~\Gamma _{x_{1}\cdots x_{k}0}. 
\]

In the IR\ limit $\lambda \rightarrow \infty ,$ the result is 
\[
\sum_{\{k_{i}\}}\prod_{i=1}^{4}\frac{1}{k_{i}!}\left( \partial _{i}^{\mu
_{i}}\right) ^{k_{i}}u(0)\int \prod_{i=1}^{k}{\rm d}^{4}x_{i}~\left(
x_{i}^{\mu _{i}}\right) ^{k_{i}}~\Gamma _{x_{1}\cdots x_{k}0}. 
\]
Here the expression $\left( x_{i}^{\mu _{i}}\right) ^{k_{i}}\left( \partial
_{i}^{\mu _{i}}\right) ^{k_{i}}$ stands for $\left( x_{i}\cdot \partial
_{i}\right) ^{k_{i}}$. Equation (\ref{eqw}) gives then 
\begin{equation}
\int \prod_{i=1}^{k}{\rm d}^{4}x_{i}~u(x_{1},\cdots ,x_{k})~\Gamma
_{x_{1}\cdots x_{k}0}^{{\rm IR}}=\sum_{\{k_{i}\}}\prod_{i=1}^{4}\frac{1}{%
k_{i}!}\left( \partial _{i}^{\mu _{i}}\right) ^{k_{i}}u(0)\int
\prod_{i=1}^{k}{\rm d}^{4}x_{i}~\left( x_{i}^{\mu _{i}}\right)
^{k_{i}}~\Gamma _{x_{1}\cdots x_{k}0}.  \label{master}
\end{equation}
This proves that the integrals on the right-hand side converge. Being this
true for arbitrary $u$, the logarithmic divergences of $\Gamma _{x_{1}\cdots
x_{k}0}$ resum to zero. From (\ref{none}), we conclude that the logarithmic
divergences of every $\overline{\Theta }$-correlator also vanish, after
resummation.

The UV\ limit $\lambda \rightarrow 0$ can be studied as follows. The local
part of $\Gamma_{x_1\cdots x_k}$ (terms of the form $\prod_{i=2}^k%
\partial^{k_i}\delta(x_1-x_i)$ with $\sum_{i=1}^n{k_i}=4$) is invariant
under the $\lambda$ rescaling. This contribution survives and gives terms
proportional to $\partial^4 U(0)= \partial^4 u(0)$. Instead, the non-local
part of $\Gamma_{x_1,\cdots,x_k}$ multiplies $\partial^4
U(x/\lambda)\rightarrow\partial^4 U (\infty )=0$ and the same integral as in
(\ref{master}), which I have just proved to be convergent. We conclude from (%
\ref{eqw}) that the local part of $\Gamma_{x_1\cdots x_k}$ is just $\Gamma^{%
{\rm UV}}_{x_1\cdots x_k}$.

Equation (\ref{master}) is the master equation generating the sum rules we
are going to study.

\medskip

{\bf Sum rule for the two-point function.} With $k=1$, we can read the sum
rule for the two-point function. We have, from (\ref{pred}) and (\ref{master}%
), 
\begin{equation}
\frac{1}{4\pi ^{2}}a_{{\rm IR}}^{\prime }\Box ^{2}u(0)=\frac{1}{4!}\partial
^{\mu }\partial ^{\nu }\partial ^{\rho }\partial ^{\sigma }u(0)~\int {\rm d}%
^{4}x~x^{\mu }x^{\nu }x^{\rho }x^{\sigma }~\Gamma _{x0}=\frac{1}{192}\Box
^{2}u(0)~\int {\rm d}^{4}x~|x|^{4}~\Gamma _{x0}.  \label{summ1}
\end{equation}
In the last step, I have used the fact that $\Gamma _{x0}$ depends only on $%
|x|$. Furthermore, writing $\Gamma =\Gamma ^{{\rm UV}}+\Gamma ^{\prime }$
(omitting the irrelevant term $\Gamma _{b}$), we have 
\begin{equation}
\Gamma _{x0}=\Gamma _{x0}^{\prime }+\frac{1}{4\pi ^{2}}a_{{\rm UV}}^{\prime
}\Box ^{2}\delta (x)=-\left\langle \Theta (x)\,\Theta (0)\right\rangle +%
\frac{1}{4\pi ^{2}}a_{{\rm UV}}^{\prime }\Box ^{2}\delta (x).  \label{ust}
\end{equation}
We then recover the known sum rule for the central charge $a^{\prime }$ \cite
{athm}: 
\begin{equation}
\Delta a^{\prime }\equiv a_{{\rm UV}}^{\prime }-a_{{\rm IR}}^{\prime }=\frac{%
\pi ^{2}}{48}~\int {\rm d}^{4}x~|x|^{4}~\left\langle \Theta (x)\,\Theta
(0)\right\rangle .  \label{sum1}
\end{equation}
I recall that the $\Theta $-correlators (encoded in $\Gamma ^{\prime }$) do
not include local contributions (which are encoded in $\Gamma ^{{\rm UV}}$).

We can now repeat this procedure to extract the sum rules for the many-point
functions of $\Theta$. I begin with the three-point function.

\medskip

{\bf Sum rules for the three-point function. } Using straightforward
identities such as 
\begin{eqnarray*}
\int {\rm d}^{4}y~y^{\rho }~\Gamma _{xy0} &=&\frac{x^{\rho }}{x^{2}}\int 
{\rm d}^{4}y~\left( x\cdot y\right) ~\Gamma _{xy0},\qquad \\
\int {\rm d}^{4}y~y^{\rho }y^{\sigma }~\Gamma _{xy0} &=&\frac{1}{3x^{2}}\int 
{\rm d}^{4}y~\left[ \delta ^{\rho \sigma }\left( x^{2}y^{2}-\left( x\cdot
y\right) ^{2}\right) +\frac{x^{\rho }x^{\sigma }}{x^{2}}\left( 4\left(
x\cdot y\right) ^{2}-x^{2}y^{2}\right) \right] ~\Gamma _{xy0},
\end{eqnarray*}
we obtain 
\begin{eqnarray*}
\int {\rm d}^{4}x~{\rm d}^{4}y~u(x,y)~\Gamma _{xy0}^{{\rm IR}} &=&\frac{1}{%
576}\int {\rm d}^{4}x~{\rm d}^{4}y~\Gamma _{xy0}~\left\{ 3\left( |x|^{4}\Box
^{x}\Box ^{x}+|y|^{4}\Box ^{y}\Box ^{y}\right) \right. \\
&&+2\left[ \left( 5x^{2}y^{2}-2\left( x\cdot y\right) ^{2}\right) \Box
^{x}\Box ^{y}+2\left( 4\left( x\cdot y\right) ^{2}-x^{2}y^{2}\right) \left(
\partial ^{x}\cdot \partial ^{y}\right) ^{2}\right] \\
&&\left. +12\left( x\cdot y\right) \left( x^{2}\Box ^{x}+y^{2}\Box
^{y}\right) \partial ^{x}\cdot \partial ^{y}\right\} u(0).
\end{eqnarray*}

Comparing (\ref{master}) with (\ref{pred}), we have the formulas 
\begin{eqnarray}
\int {\rm d}^{4}x~{\rm d}^{4}y~|x|^{4}~\Gamma _{xy0} &=&\int {\rm d}^{4}x~%
{\rm d}^{4}y~|y|^{4}~\Gamma _{xy0}=0,  \label{j1} \\
\int {\rm d}^{4}x~{\rm d}^{4}y~x^{2}\left( x\cdot y\right) ~\Gamma _{xy0}
&=&\int {\rm d}^{4}x~{\rm d}^{4}y~y^{2}\left( x\cdot y\right) ~\Gamma
_{xy0}=0,  \label{j2} \\
\frac{\pi ^{2}}{48}\int {\rm d}^{4}x~{\rm d}^{4}y~x^{2}y^{2}~\Gamma _{xy0}
&=&a_{{\rm IR}}-a_{{\rm IR}}^{\prime },  \label{j3} \\
-\frac{\pi ^{2}}{24}\int {\rm d}^{4}x~{\rm d}^{4}y~\left( x\cdot y\right)
^{2}~\Gamma _{xy0} &=&a_{{\rm IR}}-a_{{\rm IR}}^{\prime }.  \label{j4}
\end{eqnarray}

As before, we can isolate the local contributions ($\Gamma^{{\rm UV}}$) from
the rest and get

\begin{eqnarray}
\int {\rm d}^{4}x~{\rm d}^{4}y~|x|^{4}~\Gamma _{xy0}^{\prime } &=&0,
\label{i1} \\
\int {\rm d}^{4}x~{\rm d}^{4}y~x^{2}\left( x\cdot y\right) ~\Gamma
_{xy0}^{\prime } &=&0,  \label{i2} \\
\frac{\pi ^{2}}{48}\int {\rm d}^{4}x~{\rm d}^{4}y~x^{2}y^{2}~\Gamma
_{xy0}^{\prime } &=&\Delta a^{\prime }-\Delta a,  \label{i3} \\
\int {\rm d}^{4}x~{\rm d}^{4}y~~\left[ x^{2}y^{2}+2\left( x\cdot y\right)
^{2}\right] ~\Gamma _{xy0}^{\prime } &=&0.  \label{i4}
\end{eqnarray}

\medskip

{\bf Sum rules for the four-point function.} The derivation of the sum rules
for correlators with more $\Theta $-insertions is entirely similar. With the
four-point function, we have the following set of identically vanishing
relations: 
\begin{equation}
\int {\rm d}^{4}x~{\rm d}^{4}y~{\rm d}^{4}z~P_{4}(x,y,z)~\Gamma _{xyz0}=0,
\label{Pid}
\end{equation}
where $P_{4}(x,y,z)$ is one of the following monomials: $|x|^{4},x^{2}\left(
x\cdot y\right) ,x^{2}y^{2},\left( x\cdot y\right) ^{2},$ and permutations.
The other sum rules for the four-point function read 
\[
\int {\rm d}^{4}x~{\rm d}^{4}y~{\rm d}^{4}z~u(x,y,z)~\Gamma _{xyz0}^{{\rm IR}%
}=\frac{1}{2}\int {\rm d}^{4}x~{\rm d}^{4}y~{\rm d}^{4}z~\Gamma
_{xyz0}~\left( x^{\mu }x^{\nu }y^{\rho }z^{\sigma }\partial _{\mu
}^{x}\partial _{\nu }^{x}\partial _{\rho }^{y}\partial _{\sigma }^{z}+{\rm %
perms}.\right) u(0). 
\]
Using 
\begin{eqnarray*}
\int {\rm d}^{4}y\,{\rm d}^{4}z\,y^{\rho }z^{\sigma }\,\Gamma _{xyz0}={\frac{%
1}{3x^{2}}}\int {\rm d}^{4}y\,{\rm d}^{4}z\,\Gamma _{xyz0} &&\left\{ \delta
^{\rho \sigma }\left[ x^{2}\left( y\cdot z\right) -\left( x\cdot y\right)
\left( x\cdot z\right) \right] \right. \\
&&\left. +{\frac{x^{\rho }x^{\sigma }}{x^{2}}}\left[ 4\left( x\cdot y\right)
\left( x\cdot z\right) -x^{2}\left( y\cdot z\right) \right] \right\} ,
\end{eqnarray*}
we finally arrive at 
\begin{eqnarray}
\Delta a^{\prime }-\Delta a &=&\frac{\pi ^{2}}{48}\int {\rm d}^{4}x~{\rm d}%
^{4}y~{\rm d}^{4}z~\left( x\cdot y\right) \left( x\cdot z\right) ~\Gamma
_{xyz0}^{\prime },  \label{i41} \\
0 &=&\int {\rm d}^{4}x~{\rm d}^{4}y~{\rm d}^{4}z~\left[ x^{2}\left( y\cdot
z\right) -\left( x\cdot y\right) \left( x\cdot z\right) \right] ~\Gamma
_{xyz0}^{\prime }.  \label{i42}
\end{eqnarray}

\medskip

{\bf Sum rules for the correlators with a higher number of insertions. }We
have 
\begin{equation}
\int {\rm d}^{4}x_{1}~\cdots ~{\rm d}^{4}x_{n}~P_{4}(x_{1},\cdots
,x_{n})~\Gamma _{x_{1}\cdots x_{n}0}=0  \label{sum2}
\end{equation}
for $n\geq 4$ and an arbitrary degree-four polynomial $P_{4}(x_{1},\cdots
,x_{n})$.

\medskip

{\bf Sum rules for }$\Delta c${\bf . }In a similar way, from (\ref{crit}) it
is possible to derive sum rules for the trace anomaly $c$, keeping the
background metric generic. The $\Delta c$ sum rules are expressed in terms
of more complicated flow integrals, of the form 
\[
\int \prod_{i=1}^{k}{\rm d}^{4}x_{i}~P_{4}^{\mu _{1}\nu _{1}\cdots \mu
_{k}\nu _{k}}(x_{1},\cdots ,x_{k})~\frac{\delta ^{(k+1)}\Gamma }{\delta
g_{\mu _{1}\nu _{1}}(x_{1})\cdots \delta g_{\mu _{k}\nu _{k}}(x_{k})~\delta
\phi (0)}, 
\]
where now $P_{4}$ is a polynomial tensor of degree 4 in the coordinates. As
usual, it is understood that the $\Gamma $-derivatives in the integrand are
evaluated in flat space. These flow integrals involve correlators containing
intertions of one trace $\Theta $ and an arbitrary number of stress tensors $%
T_{\mu \nu }$, plus their derivatives with respect to the metric, such as $%
\delta \Theta /\delta g_{\mu \nu }$ and $\delta T_{\rho \sigma }/\delta
g_{\mu \nu }$.

\medskip

\medskip {\bf Scheme independence of the sum rules. }I now show that the sum
rules are scheme independent. The sum rules are made of flow integrals of
correlators which contain insertions of the operator $\Theta $ and the $\phi 
$-derivatives of $\Theta $, as can be read in (\ref{none}). These operators
are finite. In a unitary, renormalizable quantum field theory, $\Theta $ has
the form 
\begin{equation}
\Theta =-m_{s}^{2}\varphi ^{2}-m_{f}~\overline{\psi }\psi -\frac{\beta
_{\alpha }}{4\alpha }F_{\mu \nu }^{a~2}-\beta _{\lambda }~\varphi ^{4}-\beta
_{g}~\varphi \overline{\psi }\psi ,  \label{thetaphi}
\end{equation}
up to terms proportional to the field equations. In a generic background $%
\phi $, the $\phi $-derivatives of $\Theta $ discriminate the mass terms
from the terms proportional to the beta functions. However, $%
m_{s}^{2}\varphi ^{2}$ and $m_{f}~\overline{\psi }\psi $ are themselves
finite (by definition, the renormalization constants of $m_{s}$ and $m_{f}$
compensate the renormalization constants of $\varphi ^{2}$ and $\overline{%
\psi }\psi $, respectively). Therefore, both $\Theta $ and its $\phi $%
-derivatives are finite.

Now, the correlators of gauge-invariant, finite operators are scheme
independent at distinct points, after the perturbative series is resummed.
However, the flow integrals are sensitive also to the contributions of the
coinciding points, which, in principle, can be responsible of a scheme
dependence. We have to show that the contributions of the coinciding points
do not spoil the scheme independence of the sum rules.

To have control on the scheme effects, it is convenient to work with the
effective action $\Gamma $, rather than the separate correlators. The scheme
dependence of $\Gamma $ is classified by the set of arbitrary finite local
terms that can be added to $\Gamma $. These have the same form as $\Gamma
_{a}$ and $\Gamma _{b}$ in (\ref{bobo}), with finite coefficients. As far as
the terms of type $\Gamma _{a}$ are concerned, $\int \sqrt{g}W^{2}$ is zero
on conformally flat metrics, $\int \sqrt{g}{\rm G}$ vanishes in four
dimension (because $\sqrt{g}{\rm G}$ is a total derivative) and $\int \sqrt{g%
}R^{2}$ is responsible for the scheme dependence of $a^{\prime }$. The
quantity $a^{\prime }$ can be shifted by an arbitrary constant, independent
of the energy. This ambiguity disappears in the difference $\Delta a^{\prime
}$, which however remains dependent on the flow connecting the two fixed
points. On the other hand, the terms of the type $\Gamma _{b}$ are projected
away from the sum rules, as I\ have previously remarked.

In the sum rules, the combinations $\Gamma _{x_{1}\cdots x_{k}}^{\prime }$
appear, rather than the separate correlators. The objects $\Gamma
_{x_{1}\cdots x_{k}}^{\prime }$ are combinations of correlators such that
the scheme dependences mutually cancel. Note that the ``prime'' in $\Gamma
_{x_{1}\cdots x_{k}}^{\prime }$ (see (\ref{prime})) projects away also $%
\Gamma _{a}$ and $\Gamma _{b}$. For this reason, the sum rules are scheme
independent.

In more detail, we see from (\ref{none}) that $\Gamma _{x_{1}\cdots
x_{k}}^{\prime }$ is the sum of $-\langle \overline{\Theta }(x_{1})\,\cdots 
\overline{\Theta }(x_{k})\rangle $ plus correlators containing insertions of
the $\phi $-derivatives $\overline{\Theta }_{x_{1}\cdots x_{j}}(x_{j+1})\,$.
The correlator $\langle \overline{\Theta }(x_{1})\,\cdots \overline{\Theta }%
(x_{k})\rangle $ is scheme independent at distinct points, but scheme
dependent at coinciding points. The correlators containing insertions of $%
\overline{\Theta }_{x_{1}\cdots x_{l}}(x_{l+1})$ contribute only when some
points coincide. Their scheme depencence compensates the scheme dependence
of $\langle \overline{\Theta }(x_{1})\,\cdots \overline{\Theta }%
(x_{k})\rangle $, so that the sum $\Gamma _{x_{1}\cdots x_{k}}^{\prime }$ is
everywhere scheme independent. In practical computations, it might be
necessary to treat each correlator of $\Gamma _{x_{1}\cdots x_{k}}^{\prime }$
separately. This can lead to intermediate scheme dependent expressions. It
might also be practically difficult to separate $\Gamma $ from $\Gamma
^{\prime }$, especially non-perturbatively, while there is complete control
on the (eventually resummed) perturbative calculations.

Other approaches to the study of the scheme dependence in the stress-tensor
correlators can be found in \cite{cappelli} and \cite{jack}.

\section{Flow invariants}

\setcounter{equation}{0}

The sum rules of the previous section allow us to construct flow invariants.
A\ flow invariant is a quantity defined as the integral of a correlation
function along the RG\ flow, such that its value depends only on the fixed
points of the flow. Flow invariants are useful to characterise RG flows.
Hopefully, they can make some computations easier. For example, it might be
possible to compute $a_{{\rm IR}}$ along more convenient flows connecting
the same fixed points.

We know that $a^{\prime }$ does not have an unambiguous meaning at
criticality, but $a$ has. Consequently, $\Delta a$ depends only on the end
points of the flow, while $\Delta a^{\prime }$ can depend on the particular
flow connecting the end points. Explicit calculations \cite{inv} prove that
there are models in which the flow dependence of $\Delta a^{\prime }$ is
non-trivial. In ref. \cite{inv}, flow invariance was recovered by minimising
the flow integral (\ref{sum1}) over the trajectories connecting the same
fixed points. With the knowledge gained in the present paper, we can write
other universal flow invariants, and in particular sum rules for $\Delta a$,
eliminating $\Delta a^{\prime }$ from the identities of the previous
section. Examples of $\Delta a$ sum rules, to be used in the next sections,
are $i$) the difference between (\ref{sum1}) and (\ref{i3}) and $ii$) the
difference between (\ref{sum1}) and (\ref{i41}). Recent calculations \cite
{nuovo} have verified that the flow dependence of $\Delta a^{\prime }$ does
cancel out in the sum rules for $\Delta a$. Further results supporting this
conclusion will be published soon \cite{fest}. Therefore, the sum rules for $%
\Delta a$ are non-trivial examples of flow invariants.

Other examples of flow invariants are (\ref{i1}), (\ref{i2}), (\ref{i4}), (%
\ref{Pid}) (with the appropriate $P_{4}$'s), (\ref{i42}) and (\ref{sum2}). 

I stress once again that the sum rules are completely general. They hold (at
the perturbative level and at the non-perturbative level) in every
renormalizable quantum field theory interpolating between conformal UV\ and
IR\ fixed points. The theory need not be unitary, so also the
higher-derivative theories treated in \cite{ inv} are included. In the $%
\varphi ^{4}$-theory the stress tensor admits an improvement term and
therefore an arbitrary parameter (see \cite{hathrell2}, p. 189). This
paramenter can be fixed with the minimum principle of \cite{inv,nuovo}.

The construction of this paper naturally extends to quantum field theory in
arbitrary even dimensions \cite{fest}, where renormalizable theories are
mostly non-unitary.

\section{Examples and checks}

\setcounter{equation}{0}

To check the sum rules, I\ illustrate some examples.

\medskip

{\bf Massive scalar field. }The action in external gravity is 
\[
S=\frac{1}{2}\int {\rm d}^{4}x~\sqrt{g}\left\{ g^{\mu \nu }\partial _{\mu
}\varphi ~\partial _{\nu }\varphi +\frac{1}{6}R\varphi ^{2}+m^{2}\varphi
^{2}\right\} , 
\]
fixed by the requirement that it generates the improved stress tensor at $%
m=0 $. Focusing on the conformal factor $\phi $ and eliminating a total
derivative, we can simplify the action and write 
\[
S=\frac{1}{2}\int {\rm d}^{4}x\left\{ ~\left[ \partial _{\mu }\left( \varphi
~{\rm e}^{\phi }\right) \right] ^{2}+m^{2}\varphi ^{2}{\rm e}^{4\phi
}\right\} . 
\]
This gives 
\[
\overline{\Theta }=-\frac{\delta S}{\delta \phi }=-2m^{2}\varphi ^{2}{\rm e}%
^{4\phi }+\left( \varphi ~{\rm e}^{\phi }\right) \Box \left( \varphi ~{\rm e}%
^{\phi }\right) . 
\]
To further simplify some formulas, it is useful to subtract a term
proportional to the field equations, and define 
\begin{equation}
\widetilde{\Theta }=\overline{\Theta }+\varphi \frac{\delta S}{\delta
\varphi }=-m^{2}\varphi ^{2}{\rm e}^{4\phi }.  \label{scalar}
\end{equation}
We have 
\[
\Gamma _{x}^{\prime }=-\left\langle \widetilde{\Theta }(x)\right\rangle , 
\]
since the additional term integrates by parts to zero. We take another
functional derivative: 
\[
\Gamma _{xy}^{\prime }=-\left\langle \widetilde{\Theta }(x)~\overline{\Theta 
}(y)\right\rangle -\left\langle \frac{\delta \widetilde{\Theta }(x)}{\delta
\phi (y)}\right\rangle . 
\]
In the first term, we insert $\widetilde{\Theta }$ at the place of $%
\overline{\Theta }$. The additional term $\varphi \delta S/\delta \varphi $
can be integrated by parts. We obtain 
\begin{eqnarray}
\Gamma _{xy}^{\prime } &=&-\left\langle \widetilde{\Theta }(x)~\widetilde{%
\Theta }(y)\right\rangle +\left\langle \left( \varphi (y)\frac{\delta }{%
\delta \varphi (y)}-\frac{\delta }{\delta \phi (y)}\right) \widetilde{\Theta 
}(x)\right\rangle  \nonumber \\
&=&-\left\langle \widetilde{\Theta }(x)~\widetilde{\Theta }(y)\right\rangle
-2\delta (x-y)\left\langle \widetilde{\Theta }(x)\right\rangle .
\end{eqnarray}
Iterating this procedure, we have 
\begin{eqnarray}
\Gamma _{xyz}^{\prime } &=&-\left\langle \widetilde{\Theta }(x)~\widetilde{%
\Theta }(y)~\widetilde{\Theta }(z)\right\rangle -2\delta (x-y)\left\langle 
\widetilde{\Theta }(x)~\widetilde{\Theta }(z)\right\rangle -2\delta
(y-z)\left\langle \widetilde{\Theta }(y)~\widetilde{\Theta }(x)\right\rangle
\nonumber \\
&&-2\delta (z-x)\left\langle \widetilde{\Theta }(z)~\widetilde{\Theta }%
(y)\right\rangle -4\delta (x-y)\delta (x-z)\left\langle \widetilde{\Theta }%
(x)\right\rangle ,  \label{g3}
\end{eqnarray}
and so on.

We know, from the free-field values given in the introduction, that $\Delta
a=1/360$. Moreover, (\ref{scalar}) and (\ref{sum1}) give $\Delta a^{\prime
}=\Delta c=1/120$ \cite{cea}.

I begin the checks with (\ref{i1}). The terms of $\Gamma _{xyz}^{\prime }$
containing two $\widetilde{\Theta }$-insertions give either the integral (%
\ref{sum1}) or zero. We get the prediction 
\begin{eqnarray*}
\Delta a^{\prime } &=&\frac{\pi ^{2}m^{6}}{192}\int {\rm d}^{4}x~{\rm d}%
^{4}y~|x|^{4}~\left\langle \varphi ^{2}(x)~\varphi ^{2}(y)~\varphi
^{2}(0)\right\rangle \\
&=&\frac{\pi ^{2}m^{6}}{24}\int \frac{{\rm d}^{4}p}{(2\pi )^{4}}\frac{1}{%
\left( p^{2}+m^{2}\right) ^{2}}\left( \frac{\partial ^{2}}{\partial p^{2}}%
\right) ^{2}\frac{1}{p^{2}+m^{2}}.
\end{eqnarray*}
The calculation is straightforward in momentum space and verifies the
prediction. The check of (\ref{i2}) proceeds similarly.

The sum rule (\ref{i3}) can be converted to 
\[
3\Delta a^{\prime }-\Delta a=\frac{\pi ^{2}m^{6}}{48}\int {\rm d}^{4}x~{\rm d%
}^{4}y~x^{2}y^{2}~\left\langle \varphi ^{2}(x)~\varphi ^{2}(y)~\varphi
^{2}(0)\right\rangle . 
\]
This can be verified immediately. The sum rule (\ref{i4}) becomes 
\[
3\Delta a^{\prime }+\Delta a=\frac{\pi ^{2}m^{6}}{24}\int {\rm d}^{4}x~{\rm d%
}^{4}y~\left( x\cdot y\right) ^{2}~\left\langle \varphi ^{2}(x)~\varphi
^{2}(y)~\varphi ^{2}(0)\right\rangle 
\]
and is also verified.

For the sum rules involving the four-point function, it is necessary to
differentiate (\ref{g3}) once more and then insert it into (\ref{Pid}), (\ref
{i41}) and (\ref{i42}). Using (\ref{i41}), we get, for example, 
\begin{equation}
9\Delta a^{\prime }-5\Delta a=\frac{\pi ^{2}}{96}\int {\rm d}^{4}x~{\rm d}%
^{4}y~{\rm d}^{4}z~x^{2}\left( y-z\right) ^{2}~\left\langle \widetilde{%
\Theta }(x)~\widetilde{\Theta }(y)\,\widetilde{\Theta }(z)\,\widetilde{%
\Theta }(0)\right\rangle .  \label{b1}
\end{equation}
The check of this identity, which is verified, requires a non-trivial amount
of work, always in momentum space. Similarly, using (\ref{i42}) we arrive at 
\begin{equation}
3\Delta a^{\prime }+\Delta a=\frac{\pi ^{2}}{96}\int {\rm d}^{4}x~{\rm d}%
^{4}y~{\rm d}^{4}z~\left[ x\cdot \left( y-z\right) \right] ^{2}~\left\langle 
\widetilde{\Theta }(x)~\widetilde{\Theta }(y)\,\widetilde{\Theta }(z)\,%
\widetilde{\Theta }(0)\right\rangle .  \label{b2}
\end{equation}

The check of the identities and sum rules with more $\widetilde{\Theta }$%
-insertions are left to the reader.

\medskip

{\bf Massive fermion.} We have the action 
\[
S=\int {\rm d}^{4}x~\left[ \frac{1}{2}\left( \overline{\psi }~{\rm e}^{3\phi
/2}\right) \overleftrightarrow{\partial \!\!\!\slash }\left( \psi ~{\rm e}%
^{3\phi /2}\right) +m~\overline{\psi }\psi ~{\rm e}^{4\phi }\right] . 
\]
As above, we can define a $\widetilde{\Theta }$ subtracting the field
equations from $\overline{\Theta }$: 
\begin{equation}
\widetilde{\Theta }\equiv -\frac{\delta S}{\delta \phi }+\frac{3}{2}~%
\overline{\psi }~\frac{\delta _{l}S}{\delta \overline{\psi }}+\frac{3}{2}%
\frac{\delta _{r}S}{\delta \psi }\psi =-m~\overline{\psi }\psi ~{\rm e}%
^{4\phi }.  \label{massa}
\end{equation}
We find 
\[
\Gamma _{x}^{\prime }=-\left\langle \widetilde{\Theta }(x)\right\rangle
,\qquad \qquad \Gamma _{xy}^{\prime }=-\left\langle \widetilde{\Theta }(x)~%
\widetilde{\Theta }(y)\right\rangle -\delta (x-y)\left\langle \widetilde{%
\Theta }(x)\right\rangle , 
\]
etc. For example, we find the prediction 
\[
2\Delta a^{\prime }-\Delta a={\frac{\pi ^{2}}{48}}m^{3}\int {\rm d}^{4}x\,%
{\rm d}^{4}y\,x^{2}\,y^{2}\,\left\langle \overline{\psi }\psi (x)\,\overline{%
\psi }\psi (y)\,\overline{\psi }\psi (0)\right\rangle . 
\]
As usual, the integral can be more easily calculated in momentum space, and
gives $5/72$. This agrees with the prediction, since $\Delta a=11/360$ and $%
\Delta a^{\prime }=\Delta c=1/20$ \cite{cea}. I leave the remaining checks
to the reader.

\medskip

{\bf Yang-Mills theory with massless fermions.} We have to keep the
dimension $n$ different from 4. The action in the $\phi $-background reads 
\[
S=\int {\rm d}^{4}x~\left[ \frac{1}{4\alpha }F_{\mu \nu }^{a~2}~{\rm e}%
^{-\varepsilon \phi }+\frac{1}{2}\left( \overline{\psi }^{\,i}~{\rm e}%
^{(3-\varepsilon )\phi /2}\right) \overleftrightarrow{D\!\!\!\!\slash}%
_{ij}\left( \psi ^{j}~{\rm e}^{(3-\varepsilon )\phi /2}\right) \right] . 
\]
All quantities are bare. We define 
\begin{equation}
\widetilde{\Theta }\equiv -\frac{\delta S}{\delta \phi }+\frac{3-\varepsilon 
}{2}~\overline{\psi }~\frac{\delta _{l}S}{\delta \overline{\psi }}+\frac{%
3-\varepsilon }{2}\frac{\delta _{r}S}{\delta \psi }\psi =\frac{\varepsilon }{%
4\alpha }F_{\mu \nu }^{a~2}~{\rm e}^{-\varepsilon \phi }  \label{mamm}
\end{equation}
and get 
\begin{equation}
\Gamma _{x}^{\prime }=-\left\langle \widetilde{\Theta }(x)\right\rangle
,\qquad \qquad \Gamma _{xy}^{\prime }=-\left\langle \widetilde{\Theta }(x)~%
\widetilde{\Theta }(y)\right\rangle +\varepsilon \delta (x-y)\left\langle 
\widetilde{\Theta }(x)\right\rangle ,  \label{epsilon}
\end{equation}
etc. The second term in $\Gamma _{xy}^{\prime }$ is negligible, because it
is multiplied by $\varepsilon $. Similar terms are negligible in $\Gamma
_{x_{1}\cdots x_{k}}^{\prime }$.

The proof that these evanescent contact terms are negligible can be done as
follows. Let us consider, for example, the term $\varepsilon \delta
(x_{1}-x_{2})\left\langle \widetilde{\Theta }(x_{2})\cdots \widetilde{\Theta 
}(x_{k})\right\rangle $ in $\Gamma _{x_{1}\cdots x_{k}}^{\prime }$. Inserted
into a flow integral for the $k^{{\rm th}}$-point function of $\widetilde{%
\Theta}$, this term, after integration over $x_{1}$, produces a convergent
flow integral for the $(k-1)^{{\rm th}}$-point function, multiplied by $%
\varepsilon $. Clearly, the $\varepsilon $-factor kills this contribution.

Concluding, we can write 
\begin{equation}
\Gamma _{x_{1}\cdots x_{k}}^{\prime }=-\left\langle \widetilde{\Theta }%
(x_{1})\cdots \widetilde{\Theta }(x_{k})\right\rangle +{\cal O}(\varepsilon
).  \label{class}
\end{equation}
Similar arguments show that the contact terms of the correlators $%
\left\langle \widetilde{\Theta }(x_{1})\cdots \widetilde{\Theta }%
(x_{k})\right\rangle $ are themselves evanescent and do not contribute to
the sum rules. This can be seen from the operator-product expansion of two $%
\Theta $s.

At the level of renormalized operators, I\ recall that, in flat space, \cite
{hathrell} 
\[
\widetilde{\Theta }=-\frac{\widehat{\beta }(\alpha )}{4\alpha }\left[ F_{\mu
\nu }^{a~2}\right] +\frac{1}{2}\gamma [\overline{\psi }\overleftrightarrow{%
D\!\!\!\!\slash}\psi ]\equiv \widetilde{\Theta }^{\prime }+\frac{1}{2}\gamma
[\overline{\psi }\overleftrightarrow{D\!\!\!\!\slash}\psi ], 
\]
$\gamma $ denoting the anomalous dimension of the fermions. Here the beta
function is defined as $\widehat{\beta }={\rm d}\ln \alpha /{\rm d}\ln \mu
=\beta -\varepsilon $.

In the correlator (\ref{class}), we can freely replace $\widetilde{\Theta }$
with $\widetilde{\Theta }^{\prime }$, since the term proportional to the
fermion-field equation gives no contribution. This can be proved as follows.
First, notice that the renormalized and bare operators $\overline{\psi }%
\overleftrightarrow{D\!\!\!\!\slash}\psi $ coincide. The insertions of $%
\overline{\psi }\overleftrightarrow{D\!\!\!\!\slash}\psi $ inside the
correlators can be integrated by parts. This generates insertions of objects
of the form 
\begin{equation}
\overline{\psi }~\frac{\delta _{l}A}{\delta \overline{\psi }}+\frac{\delta
_{r}A}{\delta \psi }\psi ,  \label{bolla}
\end{equation}
with $A$ equal to $\widetilde{\Theta }$ , $\widetilde{\Theta }^{\prime }$ or 
$\overline{\psi }\overleftrightarrow{D\!\!\!\!\slash}\psi $. For $A=%
\widetilde{\Theta }$, (\ref{bolla}) is zero. This is easily seen from the
bare expression (\ref{mamm}) of $\widetilde{\Theta }$. This fact proves also
that, for $A=\widetilde{\Theta }^{\prime }$, (\ref{bolla}) is proportional
to $\overline{\psi }\overleftrightarrow{D\!\!\!\!\slash}\psi $, times a
delta function. This is trivially true also for $A=\overline{\psi }%
\overleftrightarrow{D\!\!\!\!\slash}\psi $. So, the integration by parts of
the $\overline{\psi }\overleftrightarrow{D\!\!\!\!\slash}\psi $-insertion
returns an expression which contains at least another $\overline{\psi }%
\overleftrightarrow{D\!\!\!\!\slash}\psi $-insertion. Then, the procedure
can be iterated, with a new integration by parts. In the end, we get zero.
In conclusion, a correlator with an arbitrary number of insertions of $%
\widetilde{\Theta }^{\prime }$ and at least one insertion of $\overline{\psi 
}\overleftrightarrow{D\!\!\!\!\slash}\psi $, is equal to zero.

\medskip

Formula (\ref{class}) is valid for all classically conformal quantum field
theories with a finite stress tensor. In the $\varphi^4$-theory, where the
stress tensor admits an improvement term, the finite stress tensor can be
fixed with the method of \cite{inv}.

\section{Self-consistency of the sum rules}

\setcounter{equation}{0}

The sum rules have been obtained imposing that the $\Theta $-correlators
tend to the prescribed UV and IR limits, fixed by the Riegert action for
conformally flat metrics. The Riegert action is obtained by integrating the
trace anomaly with respect to the conformal factor. In turn, the form of the
trace anomaly is fixed by dimensional counting and general covariance.

This suggests that the sum rules are mainly consequences of the general
covariance of the gravitational embedding. In this section, I prove that it
is not so. Most sum rules are related to one another by symmetry properties
of the integrands. All vanishing sum rules follow from the property that an
integrated-trace insertion is equal to the scale derivative of the
correlator. General covariance imposes only one relation among the sum
rules. This helps clarifying which sum rules have a truly non-trivial
content.

\medskip

{\bf Equivalent polynomials.} The idea is the following. A sum rule has the
general form 
\begin{equation}
\int {\rm d}^{4}x_{1}~\cdots ~{\rm d}^{4}x_{k}~P_{4}(x_{1},\cdots
,x_{k})~\Gamma _{x_{1}\cdots x_{k}0}^{\prime}=f\Delta a+g\Delta a^{\prime },
\label{genfor}
\end{equation}
with $(f,g)=(-1,1),$ $(0,1)$ or $(0,0)$. The flow integral of (\ref{genfor})
can be rewritten as 
\[
\lim_{V\rightarrow \infty }{\frac{1}{V}}\int_{V}{\rm d}^{4}x_{1}~\cdots ~%
{\rm d}^{4}x_{k}~{\rm d}^{4}x_{k+1}~P_{4}(x_{1}-x_{k+1},\cdots
,x_{k}-x_{k+1})~\Gamma _{x_{1}\cdots x_{k}x_{k+1}}^{\prime }, 
\]
where the integrals are restricted to a finite volume $V$. Using
translational invariance of $\Gamma _{x_{1}\cdots x_{k}x_{k+1}}^{\prime },$
we can set $x_{k+1}=0$ in the integrand. The $x_{k+1}$-integration
factorizes and, in the limit $V\rightarrow \infty $, it simplifies the
factor $1/V$. Expression (\ref{genfor}) is therefore recovered.

Translational invariance can be used to set any of the coordinates $x_{i}$
to zero. This generates equivalent sum rules. We can define an equivalence
relation $\sim $ among the polynomials $P_{4}$: 
\begin{equation}
P_{4}(x_{1},\cdots ,x_{k})\sim P_{4}(x_{1}-x_{i},\cdots
,x_{i-1}-x_{i},-x_{i},x_{i+1}-x_{i},\cdots ,x_{k}-x_{i})  \label{equ}
\end{equation}
for every $i$. Eq. (\ref{equ}) is found setting $x_i=0$ and renaming $%
x_{k+1} $ as $x_i$. Of course, equivalent polynomials are also those
obtained from $P_{4}(x_{1},\cdots ,x_{k})$ by permuting $x_{1},\cdots ,x_{k}$%
.

\medskip

{\bf The case $k=2$.} For example, if we take $k=2$, we get 
\begin{eqnarray*}
\left( x\cdot y\right) ^{2}~ &\sim &~\left[ (x-y)\cdot y\right] ^{2},\qquad
|x|^{4}~\sim ~|x-y|^{4},\qquad x^{2}y^{2}~\sim ~(x-y)^{2}y^{2}, \\
x^{2}\left( x\cdot y\right) ~ &\sim &~(x-y)^{2}\,(y-x)\cdot y~\sim
~y^{2}\,x\cdot (x-y)~.
\end{eqnarray*}
Working out these equivalence relations in more detail, it is easily seen
that they reduce to two independent relations, namely 
\begin{equation}
3|x|^{4}~\sim ~6x^{2}\left( x\cdot y\right) ~\sim ~4\left( x\cdot y\right)
^{2}+2x^{2}y^{2}.  \label{eqr}
\end{equation}
Therefore, the vanishing sum rules (\ref{i1}), (\ref{i2}) and (\ref{i4}) are
equivalent. Concluding, there exists a unique vanishing sum rule for the
three-point function of $\Theta$, say (\ref{i1}), generated by $%
P_{4}(x,y)=|x|^{4}$.

Observe that since $P_{4}(x,y)=|x|^{4}$ does not depend on $y$, in (\ref{i1}%
) we can single out the insertion of an integrated trace. The sum rule (\ref
{i1}) follows from the properties of this insertion, which I now derive in
generality. For simplicity, I assume that the super-renormalizable
parameters in the theory are just the masses. The derivation can be
immediately extended to theories with other super-renormalizable parameters.

\medskip

{\bf The generator of vanishing sum rules}. If $S=\int {\rm d}^{n}x\,{\cal L}
$ is the action and ${\cal L}$ the lagrangian, we define 
\[
\overline{\Theta }=-{\frac{\delta S}{\delta \phi }},\qquad \widetilde{\Theta 
}=-{\frac{\tilde{\delta}S}{\tilde{\delta}\phi }},\qquad \widehat{\Theta }=-{%
\frac{\hat{\delta}S}{\hat{\delta}\phi }}, 
\]
where 
\begin{eqnarray*}
{\frac{\tilde{\delta}A}{\tilde{\delta}\phi }} &\equiv &{\frac{\delta A}{%
\delta \phi }}-\left( 1-{\frac{\varepsilon }{2}}\right) \sum_{\varphi
}\varphi {\frac{\delta S}{\delta \varphi }}-\frac{3-\varepsilon }{2}%
\sum_{\psi }\left( \overline{\psi }~\frac{\delta _{l}A}{\delta \overline{%
\psi }}+\frac{\delta _{r}A}{\delta \psi }\psi \right) , \\
{\frac{\hat{\delta}A(x)}{\hat{\delta}\phi (y)}} &\equiv &{\frac{\tilde{\delta%
}A(x)}{\tilde{\delta}\phi (y)}}-\delta (x-y)\sum_{m}m{\frac{\partial A(x)}{%
\partial m}}.
\end{eqnarray*}
Here, $\varphi $, $\psi $ and $m$ denote collectively the scalar fields,
fermions and masses of the theory. As illustrated in the previous section, $%
\overline{\Theta }$ contains three kinds of terms: field equations of the
scalars and fermions; mass operators; evanescent terms. The operator $%
\widetilde{\Theta }$ contains only mass terms and evanescent terms. The
operator $\widehat{\Theta }$ contains just the evanescent terms.

Integrating 
\[
\Gamma _{x}^{\prime }={\frac{\delta \Gamma ^{\prime }}{\delta \phi (x)}}%
^{{}}=-\left\langle \widetilde{\Theta }(x)\right\rangle =-\left\langle 
\widehat{\Theta }(x)\right\rangle +\sum_{m}m\left\langle {\frac{\partial 
{\cal L}(x)}{\partial m}}\right\rangle 
\]
over $x$, we obtain 
\begin{equation}
\int {\rm d}^{n}x\,\Gamma _{x}^{\prime }=-\left\langle \int {\rm d}^{n}x\,%
\widehat{\Theta }(x)\right\rangle +\sum_{m}m\left\langle {\frac{\partial S}{%
\partial m}}\right\rangle .  \label{pio}
\end{equation}
A classical argument \cite{adler,nielsen}, which applies unchanged at $\phi
\neq 0$, can express the insertion of the integrated $\widehat{\Theta }$ in
terms of a $\mu $-derivative, namely 
\begin{equation}
\left\langle \int {\rm d}^{n}x\,\widehat{\Theta }(x)\right\rangle =-\mu {%
\frac{\partial \Gamma ^{\prime }}{\partial \mu }}.  \label{nl}
\end{equation}
This is proved observing that, at the bare level, $\Gamma $ depends on $\mu $
only because the bare coupling (say, the gauge coupling $\alpha $) is
dimensionful away from 4 dimensions. For a gauge coupling, we have $\alpha
=\alpha ^{\prime }\mu ^{\varepsilon }$, with $\alpha ^{\prime }$
dimensionless. Then, the $\mu $-derivative of $\Gamma $ can be written as 
\[
\mu {\frac{\partial \Gamma ^{\prime }}{\partial \mu }}=\varepsilon \alpha {%
\frac{\partial \Gamma ^{\prime }}{\partial \alpha }}=-\left\langle \int {\rm %
d}^{n}x\,{\frac{\varepsilon }{4\alpha }}F^{2}\,{\rm e}^{-\varepsilon \phi
}\right\rangle =-\left\langle \int {\rm d}^{n}x\,\widehat{\Theta }%
(x)\right\rangle . 
\]
It is easy to extend the calculation to $\varphi ^{4}$-interactions and
Yukawa couplings. Then, we can rewrite (\ref{pio}) as 
\begin{equation}
\int {\rm d}^{n}x\,\Gamma _{x}^{\prime }=\sum_{m}m{\frac{\partial \Gamma
^{\prime }}{\partial m}}+\mu {\frac{\partial \Gamma ^{\prime }}{\partial \mu 
}}.  \label{gogo}
\end{equation}
This is the generator of all vanishing sum rules.

\medskip

{\bf Vanishing sum rules with insertions of an integrated trace.} Taking $k$ 
$\phi $-derivatives of (\ref{gogo}), we get 
\[
\int {\rm d}^{n}x\,\Gamma _{xx_{1}\cdots x_{k}}^{\prime }=\left( \mu {\frac{%
\partial }{\partial \mu }}+\sum_{m}m{\frac{\partial }{\partial m}}\right)
\Gamma _{x_{1}\cdots x_{k}}^{\prime }. 
\]
Multiplying by an arbitrary degree-four polynomial $P_{4}(x_{1},\cdots
,x_{k-1})$, integrating over $x_{1},\cdots ,x_{k-1}$, and setting $x_{k}=0$,
we obtain 
\begin{eqnarray*}
\phantom{.} &&\int {\rm d}^{n}x\,\prod_{i=1}^{k-1}{\rm d}^{n}x_{i}%
\,P_{4}(x_{1},\cdots ,x_{k-1})\,\Gamma _{xx_{1}\cdots x_{k-1}0}^{\prime } \\
&=&\left( \mu {\frac{\partial }{\partial \mu }}+\sum_{m}m{\frac{\partial }{%
\partial m}}\right) \int \prod_{i=1}^{k-1}{\rm d}^{n}x_{i}\,P_{4}(x_{1},%
\cdots ,x_{k-1})\,\Gamma _{x_{1}\cdots x_{k-1}0}^{\prime }.
\end{eqnarray*}
The integrals are equal to linear combinations of $\Delta a$ and $\Delta
a^{\prime }$. These are dimensionless quantities and, in particular, they
are annihilated by the scale-derivative operator $\mu \,\partial /\partial
\mu +\sum_{m}m\,\partial /\partial m$~\footnote{%
Observe that it is not necessary to assume that $\Delta a^{\prime }$ is
independent of the dimensionful parameters $m$ and $\mu $. Indeed, it was
explicitly demonstrated in \cite{inv} that $\Delta a^{\prime }$ does depend
on the ratios among them.}. We conclude that 
\begin{equation}
\int {\rm d}^{n}x\,\prod_{i=1}^{k-1}{\rm d}^{n}x_{i}\,P_{4}(x_{1},\cdots
,x_{k-1})\,\Gamma _{xx_{1}\cdots x_{k-1}0}^{\prime }=0  \label{gho}
\end{equation}
for all $k$'s and all $P_{4}$'s. Since $\Gamma _{{\rm UV}}$ satisfies (\ref
{gho}), we can replace $\Gamma ^{\prime }$ with $\Gamma $ in (\ref{gho}).

Formula (\ref{gho}) implies the vanishing sum rules which contain at least
one integrated-trace insertion. The other vanishing sum rules can be
obtained from (\ref{gho}), by applying the equivalence relations (\ref{equ}%
). For $k=2$, this proves (\ref{i1}), (\ref{i2}) and (\ref{i4}).

\medskip

{\bf The cases $k=3,4$ and higher.}

For $k=3$, there are only two polynomials for which (\ref{gho}) does not
apply: these are $\left( x\cdot y\right) \left( x\cdot z\right) $ and $%
x^{2}\left( y\cdot z\right) $. Using (\ref{equ}) and (\ref{gho}), we get 
\[
\left( x\cdot y\right) \left( x\cdot z\right) ~\sim ~(x-y)\cdot
y\,(x-y)\cdot (y-z)~\sim ~x^{2}\left( y\cdot z\right) , 
\]
which proves (\ref{i42}). For $k=4$, the only polynomial for which (\ref{gho}%
) does not apply is $\left( x\cdot y\right) \left( z\cdot w\right) $. We
have 
\[
\left( x\cdot y\right) \left( z\cdot w\right) ~\sim ~-\left( x\cdot y\right)
\left( z\cdot w\right) ~\sim ~0. 
\]
For $k>4$, (\ref{gho}) applies to all polynomials.

\medskip

In conclusion, all vanishing sum rules can be derived from (\ref{gho}) and (%
\ref{equ}). Three non-vanishing sum rules are left: one for $k=2,$ one for $%
k=3$ and one for $k=4$. However, the independent quantities are just two: $%
\Delta a$ and $\Delta a^{\prime }$. The relation among the three
non-vanishing flow integrals is due to the general covariance of the
gravitational embedding.

\section{On the irreversibility of the RG flow}

\setcounter{equation}{0}

In this section, I discuss the issue of the irreversibility of the RG flow
and the possible relevance of the sum rules found here.

In \cite{athm}, I have shown that a physical principle, precisely the
statement that in unitary, classically conformal quantum field theories, the
induced action $\Gamma[\phi] $ is positive definite at every energy, if it
is positive definite at some energy, implies 
\begin{equation}
\Delta a=\Delta a^{\prime}\geq 0.  \label{magna}
\end{equation}
The physical principle was suggested by the consideration that only
divergences can be responsible for a violation. However, the very
evanescence of $\widehat{\Theta}$ makes $\Gamma[\phi] $ divergent-free.
Equality (\ref{magna}) has been checked to the fourth-loop order in
perturbation theory.

Independently of the arguments of \cite{athm}, sum rules for $\Delta a$ can
be written using the formulas derived in the present paper. For example, we
have 
\begin{eqnarray*}
\Delta a&=&-{\frac{\pi^2}{48}} \int {\rm d}^4 x\, |x|^4
\,\Gamma^\prime_{x0}- {\frac{\pi^2}{48}}\int{\rm d}^4 x\, {\rm d}^4 y\,
x^2\, y^2\, \Gamma^\prime_{xy0} \\
&=&-{\frac{\pi^2}{48}} \int {\rm d}^4 x\, |x|^4 \,\Gamma^\prime_{x0}- {\frac{%
\pi^2}{48}}\int {\rm d}^4 x\, {\rm d}^4y\, {\rm d}^4z\,\left(x\cdot y\right)
\left(x\cdot z\right)\Gamma^\prime_{xyz0}.
\end{eqnarray*}
The relations between $\Gamma_{x_1\cdot x_k}^\prime$ and the $\Theta$%
-correlators are read from (\ref{none}). In classically conformal field
theories, the formulas can be simplified using (\ref{class}).

The equality (\ref{magna}) of $\Delta a$ and $\Delta a^{\prime}$ amounts to
an additional, ``dynamical'' vanishing rule, not contained in the set of
``kinematic'' vanishing rules of section 6. For example, 
\[
\int {\rm d}^4 x\, {\rm d}^4 y\, {\rm d}^4 z \, \left(x\cdot y\right)
\left(x\cdot z\right)\left\langle \widehat{\Theta}(x)\,\widehat{\Theta}(y) \,%
\widehat{\Theta}(z)\,\widehat{\Theta}(0)\right\rangle=0, 
\]
in classically conformal theories. The evaluation of this integral in
perturbation theory appears to be non-trivial and I am forced to postpone
this to a future investigation.

\medskip

{\bf On the application of Osterwalder-Schrader positivity.} The next
question is how to apply Osterwalder-Schrader positivity \cite{oster}, which
states that the expression 
\begin{equation}
\sum_{k,m}\int f_k(x_1,\cdots, x_k) \,\bar{f}_m(\theta y_1,\cdots,\theta
y_m)\,\left\langle \widehat{\Theta}(x_1) \cdots\widehat{\Theta}(x_k)\,%
\widehat{\Theta}(y_1) \cdots\widehat{\Theta}(y_k)\right\rangle,  \label{legi}
\end{equation}
is non-negative, for every set of functions $f_k(x_1,\cdots, x_k)$,
vanishing unless $x_{1}^0>\cdots >x_k^0>0$. The integral is in ${\rm d}%
^4x_1\cdots{\rm d}^4x_k\,{\rm d}^4y_1\cdots {\rm d}^4y_m$ and $\theta
(x^{0},x^{1},x^{2},x^{3})=(-x^{0},x^{1},x^{2},x^{3})$. The positivity
condition holds for every choice of the $x^0$-axis.

Application of OS positivity to flow integrals of the four-point function is
not straightforward. Here I discuss the main difficulties.

The conditions on the functions $f$ exclude the coincident points of the
correlators from the integrals. The correlators might contain contact,
semi-local terms (the local terms, instead, are contained in $\Gamma^{{\rm UV%
}}$). The semi-local terms are cut away from the OS condition, but
contribute to the sum rules of the previous sections.

It is possible to write equivalent sum rules for $\Delta a$, such that the
polynomial appearing in the flow integral of the four-point function is
positive. This, however, is not sufficient to apply OS positivity, whose
formulation is intrinsically non covariant, because of the choice of a
``time'' axis and the condition that the functions $f_k$ vanish when some of
their arguments have non-positive ``times''.

I conclude that there does not appear to be a straightforward way to apply
OS positivity and prove the irreversibility of the RG flow from the
kinematic sum rules of this paper.

\section{Conclusions}

I\ have derived general sum rules for the anomalies $a$ and $a^{\prime }$,
expressing the differences $\Delta a=a_{{\rm UV}}-a_{{\rm IR}}$ and $\Delta
a^{\prime }=a_{{\rm UV}}^{\prime }-a_{{\rm IR}}^{\prime }$ as multi-flow
integrals of correlators containing insertions of the trace of the stress
tensor. Universal flow invariants are constructed by eliminating $\Delta
a^{\prime }$. The sum rules hold in the most general renormalizable quantum
field theory (unitary or not), interpolating between UV\ and IR\ conformal
fixed points. All vanishing sum rules can be derived from simple symmetry
properties, combined with the fact that an integrated-trace insertion is
equal to a scale derivative. The statements of \cite{athm} can be collected
into a further, ``dynamical'' vanishing sum rule, not contained in the set
of ``kinematic'' formulas found here. Application of Osterwalder-Shrader
positivity to flow integrals of the four-point function is not immediate.
The approach developed here can be naturally generalized to write sum rules
for $\Delta c$.

\medskip {\bf Acknowledgements} \vskip .2truecm

I am grateful to U. Aglietti, E. D'Emilio, M. Grisaru, K. Konishi, P.
Menotti, M. Mintchev, R. Rattazzi, G.C. Rossi and F. Strocchi for useful
discussions. I also thank CERN for hospitality during the final stage of
this work.

\end{document}